\newsavebox{\measurebox}
\documentclass{iau} 
\usepackage{graphicx}
\usepackage{caption}
\usepackage{adjustbox}
\usepackage{amsfonts}
\usepackage[font=scriptsize,labelfont=scriptsize]{caption}
\title[Magnetic properties of coronal flux ropes associated with CMEs] 
{ Geometric and magnetic properties of coronal flux ropes associated with CMEs
leading to geomagnetic storms}

\author[Ranadeep Sarkar, Nandita Srivastava]   
{Ranadeep Sarkar$^1$, Nandita Srivastava$^1$}

\affiliation{$^1$Udaipur Solar Observatory, Physical Research Laboratory,\\Badi Road, Udaipur 313001, India
 \\ email: {\tt ranadeep@prl.res.in} \\ email: {\tt nandita@prl.res.in}} 
\pubyear{2018}
\volume{340}  
\jname{Long-term datasets for the understanding of solar and stellar magnetic cycles}
\editors{A.C. Editor, B.D. Editor \& C.E. Editor, eds.}
\begin{document}

\maketitle

\begin{abstract}
We have studied three Interplanetary Coronal Mass Ejections (ICMEs) having clear signatures of magnetic cloud (MC) arrival at 1 AU and their associated solar sources during 2011 to 2013. Comparing the axial magnetic field strength (B$_0$) of the near-Sun coronal flux-ropes with that of the MC at 1 AU, we have found that the average inferred value of B$_0$ at 1 AU assuming the self-similar expansion of the flux-rope is two times smaller than the value of B$_0$ obtained from the results of MC fitting. Furthermore, by comparing the initial orientation of the flux-rope near the Sun and its final orientation at 1 AU we have found that the three CMEs exhibited more than 80$^\circ$ rotation during its propagation through the interplanetary medium. Our study suggests that although the near-Sun magnetic properties of coronal flux-ropes can be used to infer the field strength of the associated MC at 1 AU, it is difficult to estimate the final orientation of the MC axis in order to predict the geo-effectiveness of the ICMEs.
 
\end{abstract}

 
\firstsection 
\section{Introduction}
Predicting the strength and orientation of the magnetic field associated with the interplanetary coronal mass ejections (ICMEs) is one of the key challenges in space weather physics. Study of both the remote-sensing and in-situ data are required in order to infer the magnetic properties of ICMEs in advance. Recent studies reveal that the magnetic properties of near-Sun coronal flux ropes are well-correlated with that of the ICMEs at 1 AU (Gopalswamy et al., 2017). In this work, we aimed to study the geometric and magnetic properties of near-Sun coronal flux-ropes and that of ICMEs in order to understand the geo-effectiveness of the associated magnetic clouds.

 \begin{table}[b]
\caption {Comparison of the magnetic properties of near-Sun coronal flux ropes with that of the MCs at 1 AU } 
\begin{center}
\begin{adjustbox}{max width=\textwidth}

\begin{tabular}{cccccccccccccc} 
  \hline
 \hline
\noalign{\smallskip} 
CME                       & Tilt  & Tilt  &  \multicolumn{2}{c}{B$_0$ [mG]}&\multicolumn{2}{c}{$\phi_p$ at 10 R{$_s$}} &\multicolumn{2}{c}{$\phi_t$ at 10 R{$_s$}} & Tilt  &   B$_0$	 & $\phi_p$ & $\phi_t$&Helicity   \\
in LASCO			   & angle on  & angle from& \multicolumn{2}{c}{at 10 R$_s$} &\multicolumn{2}{c}{[$\times10^{21}$ Mx]} &\multicolumn{2}{c}{[$\times10^{21}$ Mx]}& angle of   &   [nT] at          &[$\times10^{21}$ Mx]  &[$\times10^{21}$ Mx]&[Mx$^2$] of \\
 \cline{4-9}\noalign{\smallskip} 

dd/mm/yy&solar disk	& GCS fitting&	PEA&	FRA&PEA	&FRA	&PEA	&FRA& the MC 	&1 AU	& at 1 AU & at 1 AU&the MC\\ \noalign{\smallskip} 
\hline 
\hline \noalign{\smallskip} 
01/19/2012                         &   74$^\circ$   & 60$^\circ$  & 46.9   & 32.2 &9.4	&6.4	&12.8	&8.8	&    -40$^\circ$  &    9    & 0.9&0.1 & 9$\times10^{40}$ \\
15:12 UT\\
\hline
07/12/2012                         &   45$^\circ$    & 52$^\circ$ & 130     & 48  &	1.4 &	5.3&0.6	&2.2	& -62$^\circ$   & 43   & 29.6 & 32.1& 9.5$\times10^{44}$               \\
16:48 UT\\
\hline
03/15/2013                         &   53$^\circ$    & 70$^\circ$ &40      &   --   & 3.8	&--	&1.9	&--	&   -12$^\circ$ & 13 & 0.6& 0.3& 2$\times10^{41}$  \\
 07:12 UT     \\ 
\hline

\hline

\end{tabular}
\end{adjustbox}
\end{center}
\end{table}

\begin{figure}
 \centerline{\includegraphics[width=.82\textwidth,clip=]{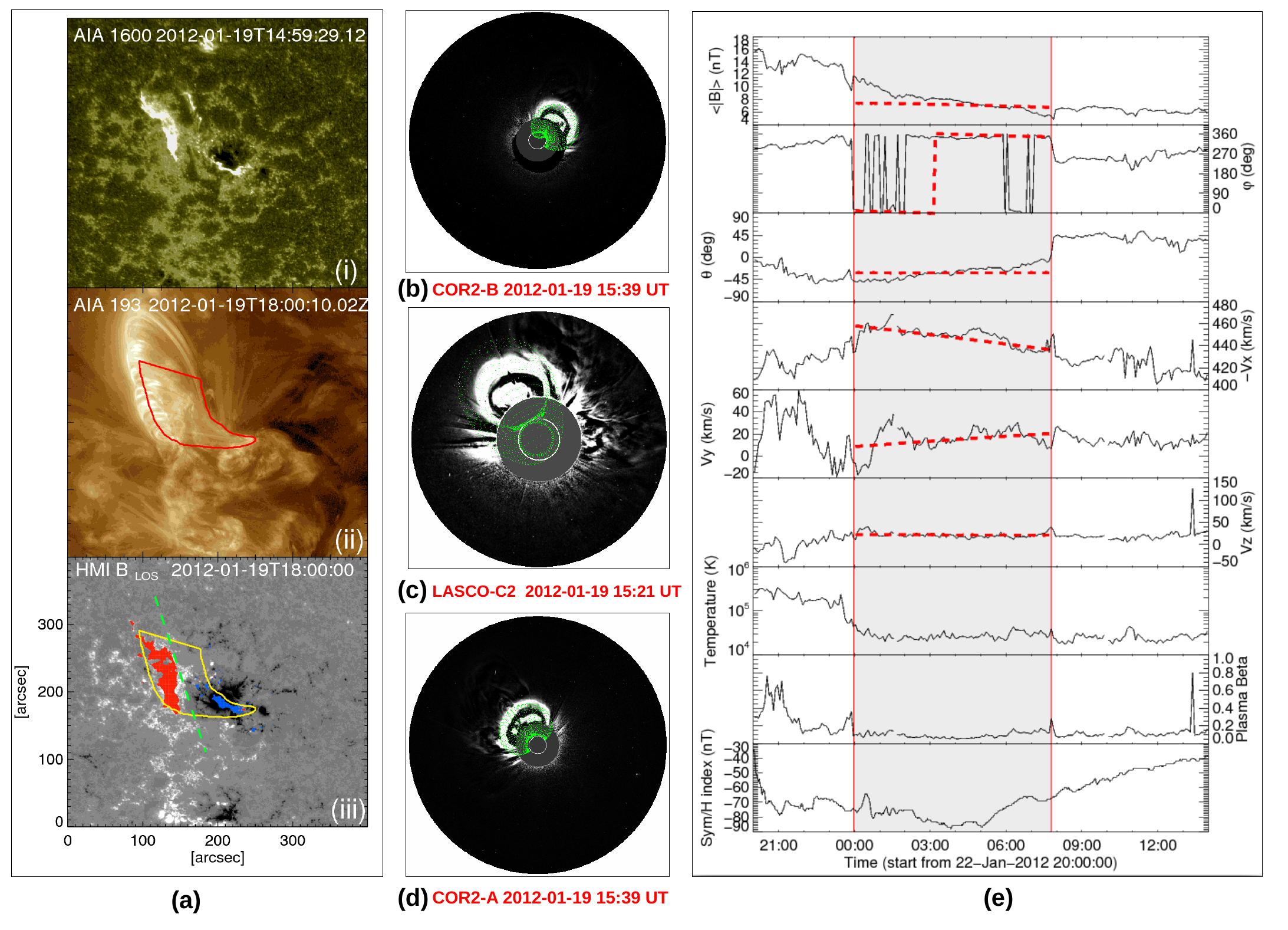}}
 \caption{\scriptsize Panel [a](i):Flare ribbons in AIA 1600 $\textrm{\AA}$ image for the M3.2-class flare on 19th January 2012.The red boundary line in Panel [a](ii) marks the PEA in AIA 193 $\textrm{\AA}$ image. Panel [a](iii): HMI line-of-sight magnetic field. The red and blue regions depict the cumulative flare ribbon area overlying the positive and negative magnetic field respectively. The yellow line mark the region underlying the PEA shown in Panel [a](ii) and the green dotted line denotes the possible orientation of the flux rope. Panels [b]-[d]: The CME morphology observed from COR2-B, LASCO C2, and COR2-A, respectively. The overplotted  green dots represent the bestfitted wireframe of the magnetic flux rope in the GCS model. Panel (e): Magnetic cloud fitting for the associated ICME.}
 \label{fig1}
 \end{figure}


\section{Summary and Discussion:}
Estimating the reconnection flux underlying the post eruptive arcades (PEA) and the cumulative flare ribbon area (FRA)(Figure 1), we have calculated the near-Sun properties of the coronal flux-ropes using the FRED model (Gopalswamy et al. 2017) for the three CMEs listed in Table 1. B$_0$ of the coronal flux ropes at 10 R$_S$ for these events ranges between 32 to 130 mG for both the PEA and FRA methods, whereas using the MC fitting method (Wang et al. 2016) we have found that B$_0$ of the associated MCs at 1 AU ranges between 9 to 43 nT. The average observed value of B$_0$ for the MCs at 1 AU is $\approx$ 22nT for these three events. This value is almost two times larger than the average inferred value of B$_0$ ($\approx$ 13 nT) at 1 AU assuming the self-similar expansion of the CMEs. Comparing the orientation of the flux-rope axis we have found that the MCs at 1 AU show 80 to 110$^\circ$ deviation from the initial orientation near the Sun. This significant deviation in the axis orientation of the MCs suggests that the interplanetary deflection and rotation of the CMEs need to be quantified in order to predict the geo-effectiveness of the associated ICMEs.   

{{\textbf{Acknowledgements} We sincerely thank Dr. Yuming Wang, USTC, China for providing the numerical
codes to fit the MC structure from the in-situ data.}}

\end{document}